%Paper: hep-lat/9508011
%From: Masahiro_Imachi <imac1scp@mbox.nc.kyushu-u.ac.jp>
%Date: Wed, 9 Aug 1995 14:46:41 +0900

%%%%%%%%%%%%%%%%%%%%%%%%%%%%%%%%%%%%%%%%%%%%%%%%%%%%%%%%%%%%%%%%%%%%%%%%%%%%%
%%%%%%%%%%%%%%%%%%%%                                        %%%%%%%%%%%%%%%%%
%%%%%%%%%%%%%%%%%%%%        This is the TEMPLETE.TEX.       %%%%%%%%%%%%%%%%%
%%%%%%%%%%%%%%%%%%%%   Copy this file, rename, and fill!!   %%%%%%%%%%%%%%%%%
%%%%%%%%%%%%%%%%%%%%                                        %%%%%%%%%%%%%%%%%
%%%%%%%%%%%%%%%%%%%%%%%%%%%%%%%%%%%%%%%%%%%%%%%%%%%%%%%%%%%%%%%%%%%%%%%%%%%%%
%\input SAKO-KUMACS
\input harvmac
%%%%%%%%%%%%%%%%%%%%%%%%%%%%%%%%%%%%%%%%%%%%%%%%%%%%%%%%%%%%%%%%%%%%%%%%%%%%
%%%%%%%%%%%%%%%%%%%%%%%%%%%%%%%%% journals %%%%%%%%%%%%%%%%%%%%%%%%%%%%%%%%
\def\AP#1{Ann.\ Phys. {\bf{#1}}}

\def\JPA#1{J. \ Phys.{\bf A{#1}}}
\def\MPL#1{Mod.\ Phys.\ Lett. {\bf A{#1}}}

\def\NP#1{Nucl.\ Phys. {\bf B{#1}}}
\def\PL#1{Phys.\ Lett. {\bf B{#1}}}
\def\PR#1{Phys.\ Rev. {\bf {#1}}}
\def\PR#1{Phys.\ Rev. {\bf {#1}}}
\def\PRB#1{Phys.\ Rev. {\bf B{#1}}}

\def\PRL#1{Phys.\ Rev.\ Lett. {\bf {#1}}}
\def\PTP#1{Prog.\ Theor.\ Phys. {\bf {#1}}}

\def \SPJ{Sov.Phys. JETP}

\def\SPJ#1{Sov.\ Phys.\  JETP {\bf {#1}}}
%%%%%%%%%%%%%%%%%%%%%%%%%%%%%%%%%%%%%%%%%%%%%%%%%%%%%%%%%%%%%%%%%%%%%

\def \PD {preprint DESY 94-229, HLRZ 94-63}
\def \PU {preprint UCLA/95/TEP/14, hep-th/9505005}
%%%%%%%%%%%%%%%%%%%%%
%%%%%%%%%%%%%%%%%underline

\def \Rn#1 {\uppercase\expandafter{\romannumeral#1}}
%%%%%%%%%%%%%%%%%%%%%%%%%%%%
%%%%%%%%%%%%%%%%%%%%%
\def\gtsim{\mathrel{\hbox{\raise0.2ex
\hbox{$>$}\kern-0.75em\raise-0.9ex\hbox{$\sim$}}}}
\def\ltsim{\mathrel{\hbox{\raise0.2ex
\hbox{$<$}\kern-0.75em\raise-0.9ex\hbox{$\sim$}}}}
\let \q=`
%%%%%%%%%%%%%%%%%% \font\bigfa=cmr10 % scaled 1095
%%%%%%%%%%%% \font\bigfb=cmr10 % scaled 1313
%print

  %pc98preview

\baselineskip=14pt plus 2pt minus 2pt

%%%%%%%
\def \Romannumeral(#1) {\uppercase\expandafter{\romannumeral#1}}

\def \Romannumeral(#1) {\uppercase\expandafter{\romannumeral#1}}
\def \Rn(#1) {\uppercase\expandafter{\romannumeral#1}}
\def\Fig(#1){$${\overline{\underline{\rm Fig.{\ \ #1}}}}$$} %%%%%%Fig%%
\def\Tab(#1){$${\overline{\underline{
 \rm Table\ \  \uppercase\expandafter{\romannumeral#1}}}}$$} %%%%%%Table%%
%\def\Tab(#1){$${\overline{\underline{\rm Table\ {  #1 }}}}$$} %%%%%%Table%%

%%%%%%%%%%%%%%%%%%%%%%%%%%%%

%%%greek%%%%%%%%%%%%%%%%%%%%%%%%%
\def \a{\alpha}
\def \b{\beta}

\def \n{\nu}

\def \x{\xi}
\def \ze{\zeta}
\def \th{\theta}
\def \vth{\vartheta}

\def \inf {\infty}
\def \pro {\propto}
\def  \ze {\zeta}
\def \kp{\kappa}
%%%%%%%%%%%%%%%%%%%%%

\def \Ft{\tilde F}

\def \fn1 {N_{f_1}}
\def \fn2 {N_{f_2}}

\def \Romannumeral(#1) {\uppercase\expandafter{\romannumeral#1}}
\def \Rn(#1) {\uppercase\expandafter{\romannumeral#1}}

%\input SAKO-KUMACS.tex
%%%%%%%%%%%%%%%%
%\input SAKO-KUMACS
\Title{KYUSHU-HET-25, SAGA-HE-86}
{\vbox{\centerline{Character Expansion, Zeros of Partition Function}
\vskip0.2em\centerline{and}
\vskip0.2em\centerline {$\theta$-Term in U(1) Gauge Theory}}}
%\vskip42pt
\centerline{Ahmed S. HASSAN, Masahiro IMACHI
\footnote{*}{e-mail:imac1scp@mbox.nc.kyushu-u.ac.jp}
 , Norimasa
TSUZUKI\footnote{**}{e-mail:tsuz1scp@mbox.nc.kyushu-u.ac.jp}
}
\centerline{Department of Physics, Kyushu University, Fukuoka 812-81, JAPAN}
\centerline{Hiroshi YONEYAMA
\footnote{***}{e-mail:yoneyama@math.ms.saga-u.ac.jp}
}
\centerline{Department of Physics, Saga University, Saga 840, JAPAN}

\centerline{\bf Abstract}
 Character expansion developed in real space renormalization group (RSRG)
 approach is applied
 to U(1) lattice gauge theory with $\th$-term in 2 dimensions. Topological
charge distribution $P(Q)$ is shown to be of Gaussian form at any $\b$(inverse
coupling constant).
The partition function $Z(\th)$ at large volume is shown to be given by the
elliptic theta function.
It provides the information of the zeros of partition function as an analytic
function of $\ze= e^{i \th}$ ($\th$ = theta parameter). These
partition function zeros lead to the
 phase transition at $\th=\pi$.
Analytical results will be compared with the MC simulation results. In MC
 simulation,  we adopt (i)``set method" and (ii)``trial function method".
\Date{08/95}
\eject

%\input harvmac.tex
%%%bbb%%%%%%%%%%%%%%%% Body %%%%%%%%%%%%%%%%%%%%%%%%%%%%%%%%%%%%%%%%%%%%%%%%%
%%%sec1
\newsec{Introduction} % begins a new section
The numerical approach to the system with $\th$-term suffers from the
difficulty, because the Boltzmann weight is complex valued and cannot be
 adopted as the probability weight directly.
The character expansion method developed in the real space renormalization
group approach is not affected by this difficulty\ref{\MI}{A. A. Migdal,
\SPJ{42}(1976)413, 743.}\ref{\KA}{ L. P. Kadanoff, \AP{100}(1976)359.}\ref
{\BT}{K. M. Bitar, S. Gottlieb and C. K. Zachos, \PR
{26}(1982)2853, \PL{121}(1983)163.}\ref\IM{M. Imachi, S. Kawabe and H.
Yoneyama,
 \PTP{69}(1983) 221, 1005.}. It plays  quite
 powerful role to  understand the gauge systems with $\th$-term
\ref\As{A. S. Hassan,
 M. Imachi and H. Yoneyama,
 \PTP{93}(1995) 161.}\ref\KOW{T. G. Kovacs and
 J. F. Wheater, \MPL{6}(1991)2827.}\ref \KOTS { T. G. Kovacs,
 E. T. Tomboulis and Z. Schram \PU .}\ref \AES{ M. Asorey, J. G. Esteve and
J. Salas, \JPA{ 27}(1994)3707.}. The calculation is based
on the orthonormal property of the irreducible group characters. It can be
performed in the system with $\th$-term without any difficulty.\par
Moreover the existence of the $\th$-term leads to some new features\ref\Ho
{ G. $^,$t Hooft,
\NP{190}[FS3](1981)455. }\ref\CR{
J. L. Cardy and E. Rabinovici, \NP{205[FS5]}(1982)1.} \ref\C{J. L. Cardy,
\NP{205[FS5]}(1982)17.} \ref\Fra  {E. Fradkin and F. A. Shaposnik, \PRL{66}
(1991)267.}\ref
\Col {S. Coleman, \AP{101}(1976)239.} \ref\SC{G. Schierholz,`` $\th$ Vacua,
 Confinement and The Continuum Limit",
 \PD.}\ref \SC {G. Schierholz,hep-lat/9409019.} in
the nature of partition function and leads to new physical situation\Fra
which are absent in the system without $\th$-term\ref\PO { A. M. Polyakov,
\NP{120}(1977)
429. }.
\par
We present in this paper the character expansion for the U(1) gauge theory
with $\th$-term in two dimensions. The action is assumed to be the sum of
Wilson
real action with real (inverse) coupling  $\b$ and standard imaginary
$\th$-term
action with the coupling $\a = \th / 2 \pi$. The expression for the partition
function at any $\b$,
i.e., in both strong and weak coupling regions,
 will be  provided  and it leads to an expression for
 topological charge $(Q)$ distribution
$(P(Q))$ at any $\b$.\par
We obtain Gaussian distribution, $P(Q) \pro \exp (- \kp_V(\b)Q^2)$,
where $ \kp_V(\b)$ is a constant determined by the value of real coupling
constant $\b$ and the size of the volume $V$. The concrete expression
of $\kp_V(\b)$ will be given. In the weak coupling regions, it can be expressed
as
$$ \kp_V(\b) = C(\b) / V \pro \b/ V,    \eqno(1.1)$$
and $C(\b)$ approaches $ C_\inf \times \b$ in  $\b \to \inf$  limit (
$C_\inf \approx 19.7$). In the strong coupling limit,
$$ C(0) = 6. \eqno (1.2) $$
This result will be compared with numerical analysis.\par
  Here we note  about the
numerical approach\ref{\Ws}{U. -J. Wiese, \NP{318}(1989)153.} .
 As mentioned above, the Boltzmann factor for the action
composed of both real and imaginary parts can not be adopted as probability
weight. So we use the probability weight given by the real action only and
calculate the topological charge distribution $P(Q)$. Then the Fourier
series of $P(Q)e^{i\th Q}$ gives us the partition function.
\par

 The comparison of $P(Q)$ given by the character expansion method
with that by numerical simulation will be given. We see quite good agreement
between these approaches in all regions of the  real coupling $\b$.\par
In the MC approach
 to obtain $P(Q)$, there is a technical difficulty related to the $Q$
 dependence of $P(Q)$. The function  $P(Q)$ is quite rapidly decreasing
 function of $Q$  at large $|Q|$ and the numerical  simulation of  $P(Q)$ in
large range of $|Q|$, at once is impossible. The ways to escape this difficulty
are\Ws\ref\cp{A. S. Hassan,
 M. Imachi, N. Tsuzuki and H. Yoneyama,``Topological charge distribution and
 $CP^1 $ model with $\th$-term,Kyushu-HET-26,SAGA-HE-91(Aug. 1995)."},

\item {(i)} The MC simulation is performed in the sets of smaller $|Q|$ range.
This is said ``set method "\ref\Bha{G. Bhanot, S. Black, P. Carter and
R. Salvador,\PL{183}(1987) 331.}
 and afterwards the results in these sets
 are gathered by adjusting neighboring results.\par
\item{(ii)} Even if the range is decomposed into sets, the obtained weight
 $P(Q)$ still changes too rapidly in each set. So we use improved  $P(Q)$.
Improved  $P(Q)$ is given by normalizing  $P(Q)$ by appropriate trial
function of  $P(Q)$ which  is obtained by foregoing small numerical
simulation or by some theoretical prediction if we have it.\par
In all regions of $\b$, the topological charge distribution is found to be
 expressed by Gaussian form
$ P(Q) \pro \exp (- \kp_V(\b)Q^2)$ quite well and we have an interesting
analytical expression for the partition function  $  Z_V(\th)$. It is expressed
as
the third elliptic function $\vth_3$ in the large volume limit($V\rightarrow
\inf$).
  It leads to infinite series of
zeros of $Z_V(\th)$ on real negative axis of $ \ze (\equiv e^{i\th})$ plane,
  since $\vth_3$ has the form of infinite product of linear
 function of $\ze$.
The closest zero to the physical point $\th = \pi (\ze = -1 )$ suggests there
 is a first order phase transition\ref\Fis {M. E. Fisher and A. N. Berker,
 \PRB{ 26},(1982) 2507 .}\ref\Itz {C. Itzykson, R. B. Pearson and J. B. Zuber,
 \NP{ 220[FS8]} (1983),415.}
 at $\th = \pi$. The distance, ${\rm
 Im}\th=\kp_V(\b)$, of this closest zero
to $\ze = -1$ is proportional to $1 / V $ at any $\b$
\footnote{*$^)$}{Wiese obtained analytical estimate at $\b = 0$.}
 and an analytical expression for
 this distance is given as a function of $\b$.\par

% We point out the similarity of the partition
%function $Z_V(\th)$ and the 1-loop scattering amplitude of closed string
%with `` twisted boundary condition". There is a close resemblance of the role
%of the parameter $\th $ in the $U (1)$ gauge theory and the `` twist angle
%$\th$" in the 1-loop closed string amplitude (``torus").\par
\vfill
\eject

%%%%%%%%%%%%%%%%%%%%%%%%%sec2
\newsec{Character Expansion}
\item{\bf(2.1)}{\bf Character Expansion} \par
We will present the character expansion for the partition function of the U(1)
gauge theory with $\th$-term. The partition function is
$$Z_V(\th)=\int [dx_l] e^{\sum_p (-s(x_p)+i v(x_p))},$$
where  $x_l$ and $x_p (p=1,....,V)$
denotes the link and the plaquette variable, respectively.
The action is given by the Wilson real action
and standard imaginary action,
$$-s(x) + i v(x) = \b {\rm cos} x + i \a x \eqno(2.1)$$
where $x = F_{01}( F_{01}$  means the field strength),
$\  {\b = 1/ g^2\ \  }{\rm and} {\ \  \a = \th/ 2\pi}$. Single
 plaquette contribution is

$$  F_p(x)=\exp \{-s(L, x)+i v(L, x) \} \eqno(2.2)$$
The partition function for volume $V$ is
$$ Z_V(\th)= \int^{\pi}_{-\pi} [dx_l] {\cal F}(x_1,..., x_V) \eqno(2.3)$$
where $F$-function {$\cal F$} is given by
$${\cal F}(x_1,...,x_V) = \sum_{Q =- \inf}^{\inf} F (x_1,...,x_V)
\delta (x_1+...+x_V-2 \pi Q).\eqno(2.4)$$
In (2.4), $Q$ denotes the topological charge (integer). By use of
 Poisson sum formula, {$\cal F$} can
be written as
$${\cal F}(x_1,...,x_V) = \sum_{l =- \inf}^{\inf} F (x_1,...,x_V)
e^{i l (x_1+...+x_V)}\eqno(2.5)$$
The functional $ F (x_1,...,x_V)$ is a product of plaquette contributions,
$$\eqalign{
  F (x_1,...,x_V)& = F_p(x_1) F_p(x_2) ......F_p(x_V) \cr
} \eqno(2.6) $$
The character expansion for $F_p$ is
$$\eqalign{
   F_p (x) & = \sum_{m =- \inf}^{\inf} \Ft_m (\b,\a) \chi_m(x),\cr
 \chi_m(x) & = \exp(imx) , m = {\rm integer},
}\eqno(2.7)$$
where $\chi_m(x)$ denotes the irreducible character specified by  integer $m$.
Then
$$\eqalign{
{\cal F}(x_1,...,x_V) &= \sum_{l =- \inf}^{\inf}F_p(x_1) F_p(x_2)
......F_p(x_V)
                        e^{i l (x_1+...+x_V)}\cr
                     &= \sum_{l =- \inf}^{\inf}\prod_{k=1}^{V} \sum_{m_k}
                          \Ft_{m_k}(\b,\a) \chi_{m_k}(x_k) \chi_l(x_k) \cr
            &= \sum_{l =- \inf}^{\inf}\prod_{k=1}^{V} \sum_{m_k}
                       \Ft_{m_k}(\b,\a) \chi_{m_{k}+l}(x_k) \cr
&= \sum_{l =- \inf}^{\inf}\prod_{k=1}^{V} \sum_{m_k}
                       \Ft_{m_{k}-l}(\b,\a) \chi_{m_k}(x_k)
}\eqno(2.8)$$
The character expansion coefficient is\As
$$\Ft_{m}(\b,\a) = \sum_n \Ft_{n}^s(\b) \Ft_{m-n}^v(\a), \eqno(2.9)$$
where character expansion of $\exp(-s)$ and $\exp(i v)(=\exp(-i \a x))$
are respectively,

$$\eqalign{
      \Ft_{n}^s(\b)&= I_n(\b), \cr
   \Ft_{l}^v(\a)&= { \sin \pi (\a +l) \over {\pi (\a +l)}},
}\eqno(2.10)$$
where $I_n(\b)$ denotes the modified Bessel function. It satisfies the relation
$$ \Ft_{m+l}(\b,\a) = \Ft_m(\b,\a +l) \eqno(2.11)$$
Then
$$
{\cal F}(x_1,...,x_V) = \sum_{l =- \inf}^{\inf}\prod_{k=1}^{V} \sum_{m_k}
                          \Ft_{m_k}(\b,\a-l) \chi_{m_k}(x_k) \eqno(2.12)$$
Now we integrate all the inner links except for the outermost link variables.
$$\eqalign{
\int_{-\pi}^{\pi} \prod_l dx_{l={\rm inner link}}{\cal F}(x_1,...,x_V) &=
     \sum_l \sum_m {\Ft_m (\b,\a-l)}^V \chi_m(x_1+...+x_V)\cr
y&=x_1+...+x_V = {\rm outermost\hskip .2cm link \hskip .2cm  variable.}
}\eqno(2.13)$$
where the orthogonality
$$\int_{-\pi}^{\pi} {dv \over 2\pi} \chi_{m_1}(x_1+v) \chi_{m_2}( x_2-v)
= \delta_{m_1m_2} \chi_{m_1}(x_1+x_2) \eqno(2.14)$$
is repeatedly used. The partition function is
$$\eqalign{
     Z_V(\th) &= \int_{-\pi}^{\pi} {dy \over 2\pi}
     \sum_l \sum_m {\Ft_m (\b,\a-l)}^V \chi_m(y)  \cr
    &= \sum_l \sum_m (\Ft_m (\b,\a-l) )^V \delta_{m,0}\cr
    &= \sum_l (\Ft_0 (\b,\a-l) )^V ,
}\eqno(2.15)$$
where
$$\Ft_0 (\b,\a-l) = \sum_{n=-\inf}^\inf  I_n(\b) {\sin\  \pi(\a+l-n) \over
        \pi(\a+l-n)}.\eqno(2.16)$$
\bigskip

%%%%%%%%%%%%%%%%%%2.2
\item{\bf(2.2)}{\bf Topological charge distribution}

Topological charge distribution appears in
$$ Z_V(\th) = \sum_{Q=-\inf}^\inf P(Q) e^{i \th Q}, \eqno(2.17)$$
and $ P(Q)$ is
$$ P(Q) = \int_{-1/2}^{1/2} d\a Z_V(\th) e^{-i 2\pi \a Q}. \eqno(2.18)$$
Eqs.(2.15) and (2.17) lead to,
$$\eqalign{
 P(Q) &= \int_{-1/2}^{1/2} d\a \sum_{l=-\inf}^\inf (\Ft_0 (\b,\a-l) )^V
        e^{-i 2\pi \a Q} \cr
&= \int_{-\inf}^\inf {d\a} (\Ft_0 (\b,\a) )^V
        e^{-i 2\pi \a Q},
}\eqno(2.19)$$
where $\Ft_0 (\b,\a)$ is given by eq.(2.16) setting $l=0$.
This is an  exact expression  in $\b$.\par
We will evaluate (2.19) in large $V$ limit. The character coefficient $\Ft_0
(\b,\a)$ is an even function of $\a$ and peaked at $\a =0$.
In large $V$ limit, $(\Ft_0(\b,\a))^V$ is a function sharply peaked at $\a =
0$.
So we can evaluate its contribution by the Taylor expansion around $\a = 0$.
$$\eqalign{
\Ft_0 (\b,\a) &=\sum_{n=-\inf}^\inf  I_n(\b) {\sin\  \pi(\a + n) \over
        \pi(\a + n)}\cr
     &=  I_0(\b) { \sin \pi\a \over
        \pi\a } + 2 \sum_{n= 1 }^\inf  I_n(\b) {{\rm sin} \pi(\a + n) \over
        \pi(\a + n)}\cr
 &= I_0(\b) (1- {\pi^2 \over 6} \a^2) - 2 \a^2 \sum_{n= 1 }^\inf
    I_n(\b) {(-1)^n \over n^2} + O(\a^4)\cr
&\cong I_0(\b) [1 - ( {\pi^2 \over 6} + { 2 \over I_0(\b)} \sum_{n= 1 }^\inf
    I_n(\b) {(-1)^n \over n^2} ) \a^2]\cr
&\cong I_0(\b) \exp (- G (\b) \a^2),
}\eqno(2.20)$$
with
$$ G(\b) = {\pi^2 \over 6} + { 2 \over I_0(\b)} \sum_{n= 1 }^\inf
    I_n(\b) {(-1)^n \over n^2},\eqno(2.21)$$
for any $\b$.\par
Now we have
$$\eqalign{
P(Q) &= 2 \int_{0}^\inf d\a \   \exp (- G(\b) V \a^2 ) {\rm \cos}(2\pi \a Q)\cr
  &= \sqrt{\pi \over GV} \exp ({-\pi^2 \over G(\b)V} Q^2)\cr
  &= \sqrt{\pi \over GV} \exp (- \kp_V (\b) Q^2).
}\eqno(2.22)$$
The quantity $\kp_V (\b)$ is defined by
$$
\kp_V (\b) ={\pi^2 \over G(\b)V}.\eqno(2.23) $$
We define
$$ C(\b) =\kp_V (\b) V= {\pi^2 \over G(\b)}.\eqno(2.24)$$
Topological charge distribution has the Gaussian form. The coefficient
$\kp_V(\b)$
has the ``$1/V$" form for all $\b$. The coefficient $C(\b)$ is plotted
 as a function of $\b$ in Fig.1.
\Fig(1)
It is given,   in the strong coupling limit, by

$$\eqalign{
G(0) &= {\pi^2 \over 6},\cr
C(0) &=6.
}\eqno(2.25)$$
In weak coupling regions $(\b \gtsim 1)$, it seems proportional to $\b$.
The value of $C(\b) / \b$, however, changes slightly with $\b$ (Fig.2).\par
\Fig(2)
\vskip 1cm
%%%%%%%%%%%2.3
\item{\bf(2.3)}{\bf Partition  function zeros and phase transition}\par
When $P(Q)$ is given by Gaussian
form (2.22), the partition function
$$ Z_V(\th) = \sum_Q P(Q) e^{2\pi i \a Q} = \sum_{Q=-\inf}^\inf P(Q) \ze^Q,
\eqno(2.26)$$
has a simple factorized form
\footnote{*$^)$}{The authors are grateful to Prof. Y. Yamada for valuable
discussion
  on this point.}
$$\eqalign{
  Z_V(Q) &= \sqrt{ \pi \over GV}\sum_Q \exp (-\kp_V(\b) Q^2) e^{2\pi i \a
Q},\cr
  &= \sqrt{ \pi \over GV} q_0 \prod_{n=1}^\inf (1 + q^{2n-1} \ze)(1 + {
q^{2n-1} \over \ze}),
}\eqno(2.27)$$
where $\ze \equiv e^{i\th} = e^{2\pi i \a}$ and
$$\eqalign{
 q_0 &= \prod_{n=1}^\inf (1- q^{2n}),\cr
  q &= \exp(-\kp_V(\b)),\ (<1).
}\eqno(2.28)$$
It can be written as
$$ Z_V(\th) = \sqrt{ \pi \over GV} \vth_3(\nu,\tau),\eqno(2.29)$$
where $\vth_3(\nu,\tau)$ is the third elliptic theta function with
$q = \exp (i\pi \tau )\ {\rm and}\  \a = \nu$.\par

Important feature of (2.27) is that the partition function has infinite zeros
(at $V \to \inf$) on the real negative $\ze$ axis in complex $\ze$ plane.
  It shows that the system
undergoes transition at $\th = \pi$. When V is finite, the closest zero of
$Z_V(\th)$
to $\ze = -1 (\th = \pi)$ is located at the distance
 ${\rm Im}\th=\kp_V(\b)\pro 1/V$ from $\ze = -1$ point.
 In $ V \to \inf$ limit, it approaches the $\ze = -1$ point and infinitely
 many zeros accumulate on the real negative  axis of complex $\ze$ plane.\par
Poisson sum formula applied to (2.17) gives
$$\eqalign{
\sum_Q \exp (- \kp_V Q^2 &+ 2\pi i \a Q)\cr
&=\sum_m \int_{-\inf}^\inf\ d\x \exp (- \kp_V \x^2 + 2\pi i \a \x + 2\pi i \x
m)\cr
&=\sqrt{\pi \over \kp_V}\sum_m \exp (-{\pi^2 \over \kp_V}( \a-m)^2).
}\eqno(2.30)$$
Note that this is a  periodic Gaussian form and that it shows
clearly the phase structure as follows;
For $m_0-1/2<\a <m_0+ 1/2$, the sum is dominated by $m=m_0$ and for $m_0+1/2<
\a <m_0+ 3/2$ the sum
is dominated by $m=m_0+1$. It shows that there are infinitely many
 phase transitions at
$\a=m_0+1/2$, i.e.,
 at $\th = \pi+2 \pi m_0( m_0 =$ any integer). \par

Eq. (2.30) can also be written as,

$$
 \sqrt{\pi \over \kp_V} \exp ( - {\pi^2 \a^2 \over \kp_V}) \sum_m
\exp (-{\pi^2 \over \kp_V} m^2 - { 2 \pi^2 \a^2 \over \kp_V}m ).
\eqno(2.31)$$

The left hand side of (2.30) is $\vth_3(\nu,\tau)$. Eq.(2.31) is just
the same form as the left hand side of (2.30)
 if we replace $ \kp_V \to \kp_{V}' =\pi^2 / \kp_V$,
$\a \to \a' = i\a \pi / \kp_V$. Then we have
$$ \vth_3(\nu,\tau) = \sqrt{1 \over i \tau} \exp ({i \pi \nu^2 \over \tau})
\vth_3(\nu',\tau')\eqno(2.32)$$
where
$$\eqalign{
\nu &= \a, \cr
 i\pi \tau &=- \kp_V,\cr
\tau'&= -1/\tau, \cr
 \nu' &= \nu / \tau.
}\eqno(2.33)$$
Eq (2.33) is a kind of modular transformation.
\par
\vskip 1cm
%%%%%%%%%%%%%%%%%%%%% sec3
\newsec{ Comparison with numerical method}
When we adopt the action with $\th$-term, the Boltzmann factor
$$ \exp (- s(x) + iv(x) ) $$
is a complex number and cannot be directly used as a probability weight.
The partition function can be written
$$\eqalign{
 Z_V(\th) &= \int [dx_l]\exp ( - \sum_p s(x_p) + i \a \sum_p x_p) \cr
&= \sum_Q \int [dx_l]^{(Q)} \exp ( - \sum_p s(x_p)) e^{2\pi i \a Q} \cr
&= \sum_Q P(Q) e^{ 2\pi i \a Q}
}\eqno(3.1)$$
where  $[dx_l]^{(Q)}$ is the integration measure restricted to topological
charge $Q$.
We first calculate the topological charge distribution $P(Q)$
with real positive probability weight $\exp(-\sum s(x_p))$ and then we obtain
 $Z_V(\th) $
through Fourier series.
In order to perform numerical simulations to obtain $P(Q)$, a technical
problem appears because $P(Q)$ is usually a quite rapidly decreasing
function of $Q$ and it drops by many decades at large $Q$.
Two technical methods to avoid this difficulty will be used.
\item{(i)}``Set method": Wide range of $Q$ is divided  into many sets.
 Each set composed of $ [Q, Q+1, ..., Q+s-1]$. In the actual calculation
we choose $s =4 $. Neighboring sets are adjusted at the overlapping $Q$
 so as to give $P(Q)$ over these neighboring sets with
the common normalization. Consider two neighboring sets $ [Q-(s-1),\cdots, Q]$
and
$ [Q, Q+1, ..., Q+s-1]$. We equate the value of $P(Q)$ at
 largest $Q(=Q)$ of the lower set
with that at the smallest $Q(=Q)$ of the upper set. \par
\item{(ii)}``Trial function method (Improved $P(Q)$)": Even we divide into
sets, the change of $P(Q)$
in each set is still large. We define improved $\bar P(Q)$ dividing $P(Q)$ by
 trial $P(Q)$, `` $P_t(Q)$"
$$\bar P(Q) = P(Q) / P_t(Q).\eqno(3.2) $$
Eq.(3.2) will give very slowly changing or almost flat distribution
if we use appropriate trial distribution $P_t(Q)$.
We perform Monte Carlo simulation for U(1) gauge theory at $\b =0.0$
 (strong coupling limit), $ 1.0, 2.0, 3.0, 4.0 $
 and $  5.0$
(weak coupling regions), at volume $V = 100$ and 400.
\par
%%%%%%%%%%%%%%%%% table 1
%\Tab(1)
\vskip 1cm
\centerline{ Table I}
\vskip 1cm
\centerline{
\vbox{\offinterlineskip
\hrule
\halign{&\vrule#&
  \strut\quad\hfil#\quad\cr
height2pt&\omit&&\omit&&\omit&&\omit&\cr
&$\b$\hfil&&$C(\b)$\hfil&&$C(\b)/\b$\hfil&&measured $C(\b)$\hfil&\cr
&\hfil&&\hfil&&\hfil&&$V$=100\hskip 1cm$V$=400&\cr
height2pt&\omit&&\omit&&\omit&&\omit&\cr
\noalign{\hrule}
height2pt&\omit&&\omit&&\omit&&\omit&\cr
&0\hfil&&6\hfil&&$\inf$\hfil&&5.99\hskip 1.8cm 5.53&\cr
&0.5\hfil&&8.410\hfil&&16.79\hfil&&\hskip 1cm &\cr
&1.0\hfil&&12.30\hfil&&12.30\hfil&&12.4\hskip 1.8cm 12.0&\cr
&1.2\hfil&&14.38\hfil&&11.98\hfil&&\hskip 1cm &\cr
&2.0\hfil&&25.82\hfil&&12.91\hfil&&24.4\hskip 1.8cm 25.8&\cr
&2.4\hfil&&33.13\hfil&&13.80\hfil&&\hskip 1cm&\cr
&3.0\hfil&&45.20\hfil&&15.07\hfil&&40.5\hskip 1.8cm 45.1&\cr
&4.0\hfil&&66.19\hfil&&16.55\hfil&&57.8\hskip 1.8cm 66.3&\cr
&5.0\hfil&&86.87\hfil&&17.37\hfil&&\hskip 1cm 85.7&\cr
&10.0\hfil&&186.8\hfil&&18.68\hfil&&\hskip 1cm &\cr
&15.0\hfil&&285.8\hfil&&19.05\hfil&&\hskip 1cm &\cr
&20.0\hfil&&384.6\hfil&&19.23\hfil&&\hskip 1cm &\cr
&100.0\hfil&&1960\hfil&&19.60\hfil&&\hskip 1cm &\cr
height2pt&\omit&&\omit&&\omit&&\omit&\cr}
\hrule}
}
\vskip 1cm
 Measurements are made $10^4$ times
 for each set. Measured range of $Q$ are from 0 to $Q_{\rm Max}$ = 15 and
from 0 to 30 for
 $V$ = 100 (=$10 \times 10$) and 400 (=$20 \times 20$), respectively.
$P(Q)$ changes by $O(10^{-56})$ for $\b = 4.0$ at $V = 100$ and by
$O(10^{-82})$
 for $\b$ = 5.0
at $V$=400 between $Q=0$ and $Q=Q_{\rm Max}$.\par
The distributions obtained through MC simulation show Gaussian distribution of
$Q$
for all the cases. They are shown in fig.3(a) and (b).\par

\Fig({3(a)} )
\Fig({3(b)})

The data for $P(Q)$ is taken by the histogram method and the multinomial
distribution is assumed to obtain statistical error in each set. The obtained
error is $\approx$ 2\% of the measured $P(Q)$ and much smaller than the
 symbol of data points in the figure.\par

The slope $\kp_V(\b)$ of $Q^2$ in the Gaussian form coincides quite well
 with the analytical prediction of eq (2.23). The slope parameter $\kp_V(\b)$
obtained by numerical calculations
shows $1/V$ law for each $\b$ and the value $C(\b)\equiv \kp_V(\b) V$
( eq.(2.23) and (2.24)) is almost $V$-independent( Table (I)).
The fact that all the $P(Q)$'s measured for $\b $= 1.0, 2.0, 3.0, 4.0
 and 5.0 have the Gaussian
distribution  shows that the system
undergoes the first order phase transition at $\th = \pi$ for all these
$\b$ values.
%%%%%%%%%%%%%%%%%%%%%%%%%sec4
\newsec{Conclusions}
%\item{\bf(2.1)}{\bf Character Expansion} \par
We presented the analytic form of partition function $Z(\th)$
for U(1) lattice gauge theory with $\th$ -term in 2  dimensions.
Character expansion of RSRG
 can be safely applied to the system with $\th$-term. In character expansion
 method, only
orthonormal property of irreducible characters is used and the problem of
 complex Boltzmann
factor is harmless in this approach. Topological charge distribution for
any real coupling
constant $\b$ is given analytically, and is shown to have Gaussian form.
The coefficient
 $\kp_V(\b)$ in Gaussian form is analytically given. This form of $P(Q)$
leads to $Z(\th)$ expressed by
 the elliptic theta function $\vth_3$ in $ V\rightarrow \infty$   limit.
 This function
is known to have infinite numbers of zeros on the real negative axis of
$\ze=e^{i \th}$.
The closest zero to $-1$ provides the information about the deconfining
phase transition
at  $\th=\pi$. The distance from $\ze=-1$ is ${\rm Im }\th=\kp_V(\b)\pro
1/V$ indicates that the phase transition is
 consistent with the first order one\Fis\Itz.\par

The results for $P(Q)$ given above at any $\b$ is compared with the numerical
 simulations.
In numerical simulations the famous problem of complex Boltzmann factor
appears.
The  numerical results thus obtained agree with the
analytical results quite well.
Topological charge distribution $P(Q)$ obtained by numerical simulations
 obeys
Gaussian form.
The coefficient $\kp_V(\beta)$ of $Q^2$ in the exponent of the Gaussian is
 proportional to
$1/V$. We covered a wide range in both of $Q$ (up to $Q_{\rm Max}=30$)
 and $P(Q)$ (changes by
order $10^{-82}$).
We comment here on numerical simulations of $P(Q)$ of $CP^1$ model
\cp in two dimensions.
The results of the model at strong couplings also show Gaussian form,
 and the coefficient
obeys $1/V$ law.
\par

Poisson sum formula leads to an interesting expression eq.(2.30) for
the partition function. It is given by periodic Gaussian form. Since $1/\kp_V
\pro V\rightarrow {\rm large}$, this expression shows phase transition
clearly. For $m_0-1/2<\a<m_0+1/2$ (i.e. $m_0-\pi<\th<m_0+\pi$), $m=m_0$
dominates due
to large coefficient in front of $(m-\a)^2$. For $m_0+1/2< \a <m_0+ 3/2(\pi
+2 \pi m_0
 < \th < 3\pi+2 \pi m_0$),
 $m=m_0+1$ dominates. From this argument we understand the critical value $\a_c
 =m_0+1/2(m_0=$ any integer). Poisson sum formula also relates $\vth_3(\n,
\th)$ to a modular
 transformed form    $\vth_3(\n^{'}, \th^{'})$( (2.32 and (2.33)).
%%%%%%%%%%%%%%%%%%%%%%%%%sec4
%\newsec{Conclusions}
%\item{\bf(2.1)}{\bf Character Expansion} \par

%%%%%%%%%%%%%%%%%%%%% References %%%%%%%%%%%%%%%%%%%%%%%%%%%%%%%%%%%%%%%%%%%%

%%%%%%%%%%%%%%%%%%%%%ref
%\outrefs
\listrefs

%\listrefs
%%%%%%%%%%%%%%%%%%Table  captions
\leftline\bf{TABLE CAPTIONS}
\vskip 0.2cm
\item{Table I}Coefficients in Gaussian distributions are shown for $\b$ at
             0, 0.5, 1.0, $\cdots$ 100.\par
\vskip 1cm
%%%%%%%%%%%Fig captions
\leftline\bf{FIGURE CAPTIONS}
\vskip 0.2cm
\item{Fig. 1\ \ \ }  $\ln [C(\b)]$ vs.
             $\ln [\b]$.  $C(\b)$ is
             approximately linear function of $\b$ in weak coupling regions.
\par
\vskip 0.2cm
\item{Fig. 2\ \ \ }  $C(\b)/\b$ vs. $\b$. We can see approximately constant
                behavior at
              large $\b$.\par
\vskip 0.2cm
\item{Fig. 3(a)}Topological charge distribution $\log_{10} (P(Q))$ vs. $Q^2$
measured
               by Monte Carlo simulations for various $\b$'s. The value
               of $Q$ ranges from 0 to 15( =$Q_{\rm Max}$ )  and the lattice
               size $L \times L$ is 10$\times$10. %\par
               The inverse coupling constant $\b$ are taken to be
               0.0( square), 1.0( diamond), 2.0( circle), 3.0( triangle),
               4.0( cross). The lines show fittings according to
               the Gaussian form.\par
\vskip 0.2cm
\item{Fig. 3(b)}Topological charge distribution $\log_{10} (P(Q))$. Notations
are as
               in Fig. 3(a). $Q_{\rm Max}$=30 and $L=20$. Data for $\b=5.0$
               (nabla)  is added.\par
\vfill
\eject

\input epsf
\vskip 3cm
$$\vbox{
\vskip 3cm
\epsfysize=0.65\hsize
\epsfbox{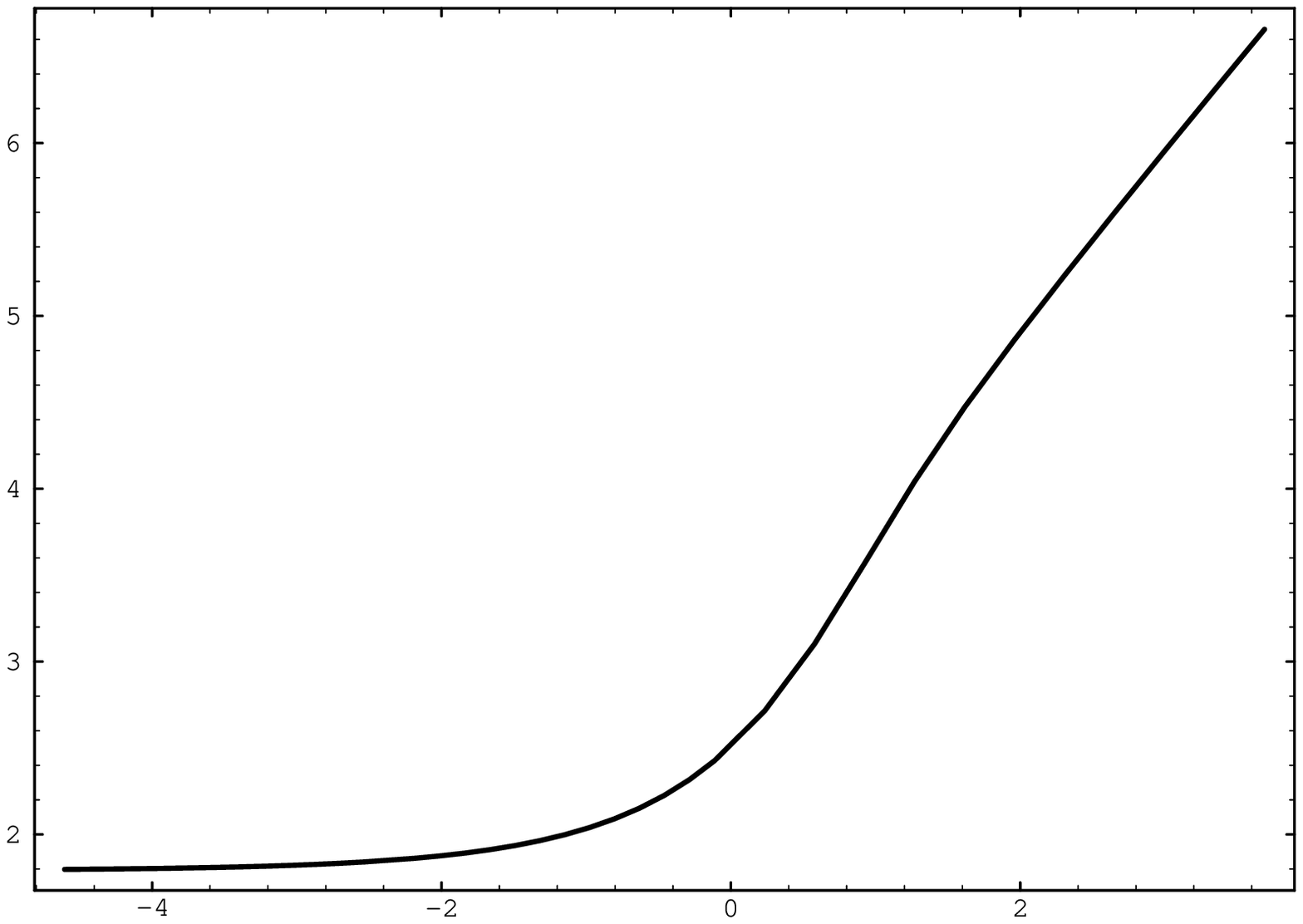}
  \centerline {Fig. 1}}$$

\vfill
\eject

$$\vbox{
\vskip 3cm
\hskip 0.5cm
\epsfysize=1.0\hsize
\epsfbox{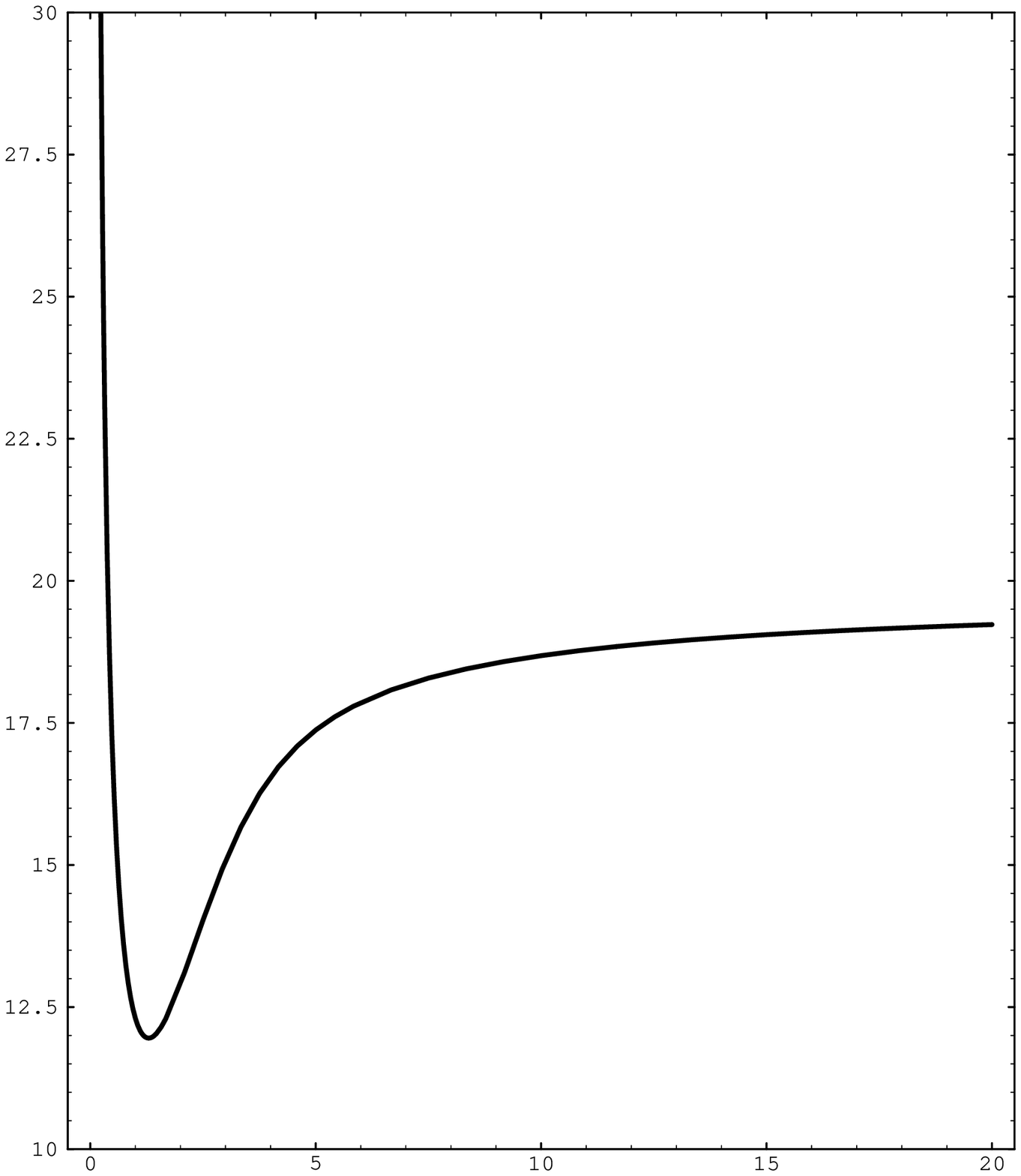}
}$$
\centerline { Fig. 2}
\vfill
\eject

\vskip 6cm
\epsfbox{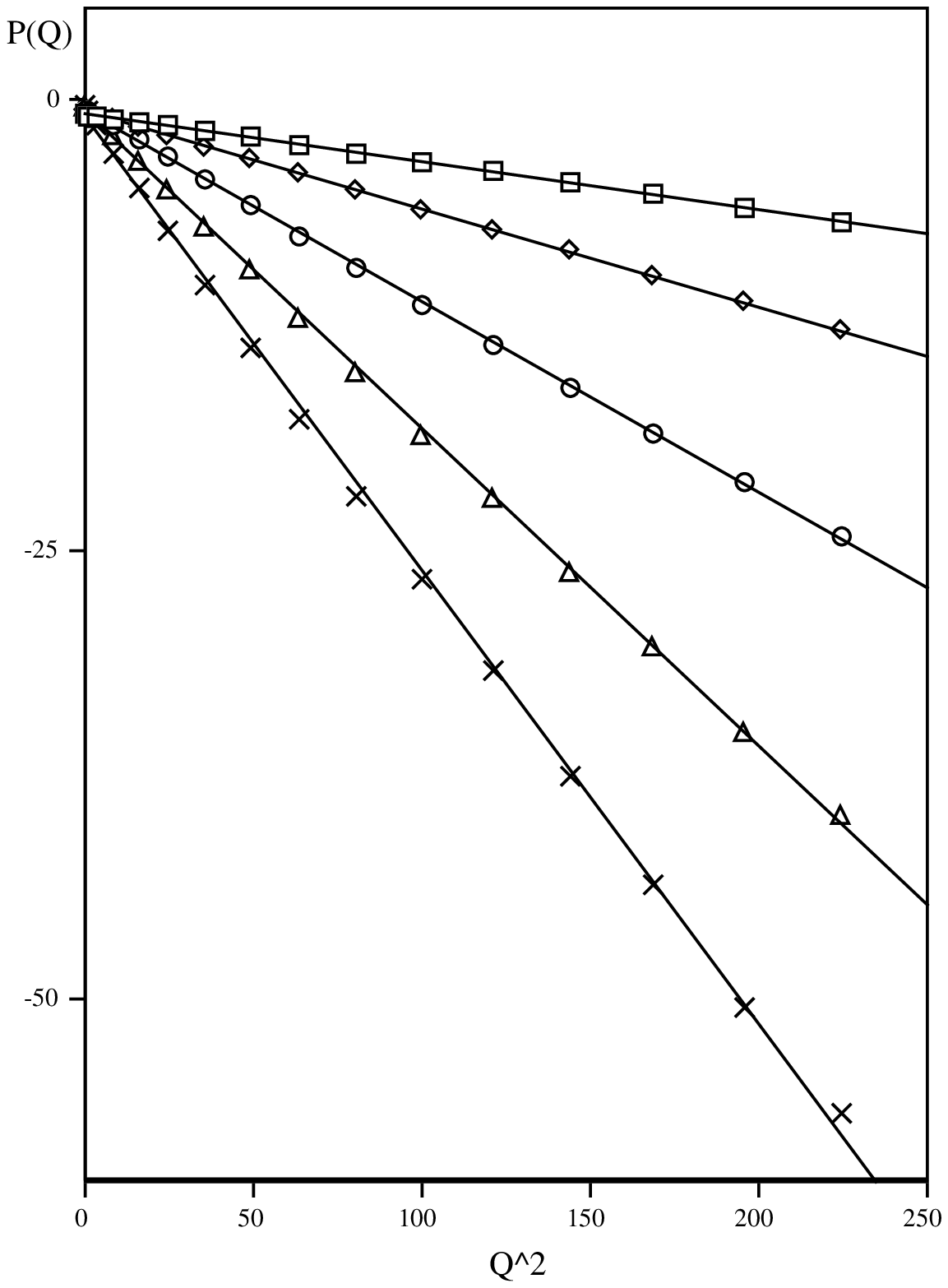}
\centerline {Fig. 3(a)}
\vfill
\eject

\vskip 6cm
\epsfbox{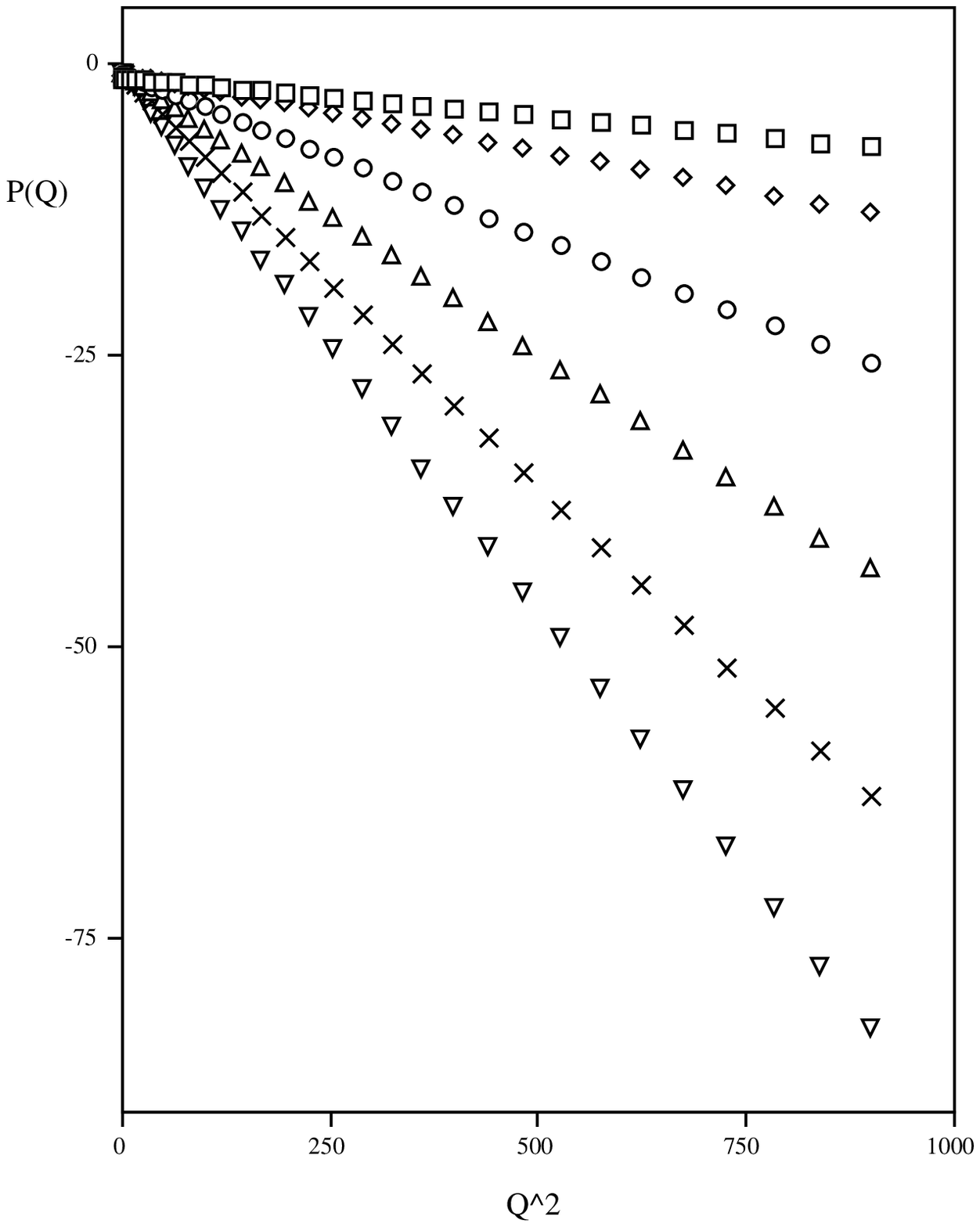}
\centerline {Fig. 3(b)}

\bye